\documentclass[twocolumn]{aa}
\usepackage{graphicx,natbib}

\def\Epk{E_{\rm p}}
\def\Ep{E_{\rm p}}
\def\Dt{\Delta t}
\def\F0{F_{\rm 0}}
\def\E00{E_{\rm 0}}
\def\Enoll{E_{\rm 0}}
\def\t0{t_{\rm 0}}

\def\e1{\eta-1}
\def\P0{\Phi_{\rm 0}}

\def\lag{\Delta t}

\def\tr{t_{\rm r}}
\def\Fm{F_{\rm m}}
\def\tm{t_{\rm m}}
\def\Em{E_{\rm m}}
\def\S{Sect. }

\newcommand{\ltsima} {\; \buildrel < \over \sim \;}
\newcommand{\gtsima} {$\; \buildrel > \over \sim \;$}
\newcommand{\lta} {\lower.5ex\hbox{\ltsima}}
\newcommand{\gta} {\lower.5ex\hbox{\gtsima}}

\begin{document}

\title{Interpretations of gamma-ray burst spectroscopy\\
I. Analytical and numerical study of spectral lags}
\author{F. Ryde}
 \offprints{F. Ryde}
 \institute{Stockholm Observatory, AlbaNova
University Center, SE-106 91 Stockholm, Sweden}
\date{Received 28 May 2004; Accepted 15 September 2004}
 \abstract{
We describe the strong spectral evolution that occurs during a
gamma-ray burst (GRB) pulse and the means by which it can be
analyzed. In particular, we discuss the change of the light curve
as a function of energy and the spectral lag. Based on observed
empirical correlations, an analytical model is constructed which
is used to describe the pulse shape and quantize the spectral lags
and their dependences on the spectral evolution parameters. Using
this model, we find that the spectral lag depends mainly on the
pulse-decay time-scale and that hard spectra (with large spectral
power-law indices $\alpha$) give the largest lags. Similarly,
large initial peak-energies, $\E00$, lead to large lags, except in
the case of very soft spectra. The hardness ratio is found to
depend only weakly on $\alpha$ and the
hardness-intensity--correlation index, $\eta$. In particular, for
low $\E00$, it is practically independent, and is determined
mainly by $\E00$. The relation between the hardness ratio and the
lags, for a certain $\E00$ are described by power-laws, as
$\alpha$ varies. These results are the consequences of the
empirical description of the spectral evolution in pulses and can
be used as a reference in analyses of observed pulses. We also
discuss the expected signatures of a sample of hard spectral
pulses (e.g. thermal or small pitch-angle synchrotron emission)
versus soft spectral pulses (e.g. optically-thin synchrotron
emission). Also the expected differences between a sample of low
energetic bursts (such as X-ray flashes) and of high energetic
bursts (classical bursts) are discussed.

   }

\authorrunning{Ryde}
\titlerunning{GRB Spectroscopy}

   \maketitle

\section{Introduction}

The analysis of the prompt emission from gamma-ray bursts (GRBs)
is giving us valuable clues to the environment from which the
radiation emanates. Much evidence points to the fact that the
gamma radiation stems from dissipation processes in a
relativistically expanding plasma wind \citep{ree92, mesrev},
either as shocks or in magnetic reconnections. Complex and
variable light curves, that consist of several, often overlapping
spikes or pulses, tell us through their variability and spectral
power distribution about the energetics \citep{kumar}, radial
structure of the fireball \citep{bel00}, and can even give hints
about the progenitor (\cite{KRM, SalGal}). Individual {\it
pulses}, which are assumed to be related to single emission
episodes, give us information on the microphysics; radiation
processes \citep{piran99}, and comoving properties of the
fireball, such as the densities and magnetic fields \citep{RP02}.
We will here review the various spectroscopic analysis methods
that are used to characterize the spectral evolution during bursts
and especially during their constituent pulses. We will examine
the analysis of the spectral lags between the light curves,
measured in different energy bands, and how these results compare
to the high-resolution spectroscopic analysis in which the
time-evolution and correlations of spectral parameters are
deduced. The latter approach is useful as it can be used directly
to test models of the dynamics and the emission processes.
However, it can only be performed on cases in which we have
high-resolution spectra and know their time-evolution. On the
other hand, the lag description has the advantage that it can be
measured on practically every GRB, even weak ones, from all
satellite missions which have more than one spectral channel.
Moreover, several relations have been identified between the
spectral lag and the physical quantities of the burst, such as the
isotropically equivalent luminosity (\cite{norris2}) and the peak
energy of the spectrum (\cite{amati}). It is therefore of interest
to understand how these two descriptions relate to each other,
since this will facilitate the interpretation of the spectral lag,
and its dependence on other parameters.

The observation that the light-curves in different energy bands
lag behind each other is a common feature in astrophysical
objects, and is not only found in GRBs. There are several physical
scenarios in which this can be explained. Positive lags, that is,
the soft radiation lagging the hard, which is the dominant
behavior observed in GRBs, can be due to an intrinsic cooling of
the radiating electrons, which will cause the radiation to
dominate at lower and lower energies. Alternatively, a Compton
reflection of a medium at a sufficient distance from the initial
hard source will also cause a lag. Also, a convex surface that
emits at relativistic speeds will cause the radiation emitted off
axis, to be delayed and softened, thus producing a lag, that is,
the curvature effect (\cite{RP02}). Furthermore, if there were an
intrinsic time-scale that is constant from burst to burst, the
change in viewing angle between the object and the observer, will
cause both the observed time-scale and intensity to vary in a
specific way due to the angular dependence of the Doppler boost
\citep{salm,nakar}. A possible scenario for negative lags, that
is, the hard radiation lagging behind the soft, is a hot medium,
say a lepton cloud surrounding a cooler emitter, which will cause
the soft radiation to be upscattered by inverse Comptonization.
The photons get harder the more scatterings they suffer and thus
they are more delayed. A definite answer to the reason for the
spectral lags in GRBs has not yet been given.

We will expand the spectral-lag analysis done sofar (denoted here
as low-resolution spectroscopy, LRS: see, e.g., \cite{band97} and
references therein) by connecting it to the detailed
spectral-evolution description that can be done for strong pulses
(denoted here as high-resolution spectroscopy, HRS: see, e.g.,
\cite{ford95,crider97,RS02}). The HRS and LRS are reviewed in \S
\ref{sec:empirical} and in \S \ref{sec:analytical} we give an
analytical treatment assuming a simple pulse model. The general
empirical behavior of bursts, found through the observations made
by the Burst and Transient Source Experiment (BATSE) on the {\it
Compton Gamma-Ray Observatory} and their parameter distributions,
are used in combination with the analytical findings to make a
numerical study in \S \ref{sec:simulations}. We simulate realistic
pulses and determine how various spectral evolutions are
manifested as spectral lags. Finally, we discuss the
interpretations that can be made in \S 5.

In a subsequent paper II (Ryde et al. 2004), we will demonstrate
the results on a sample of GRB pulses observed by BATSE and
discuss the spectral lag correlations and various model scenarios.

\section{GRB Spectral Evolution: Empirical View}
\label{sec:empirical}

\subsection{High Resolution Spectroscopy (HRS)}
\label{sec:HRS}

For the strongest BATSE pulses, spectroscopy is possible with high
resolution, allowing us to deconvolve the observed count spectra
through the detector response, thus providing the incoming photon
spectrum. The deconvolution is done through a forward-fitting
method. This is done for several time intervals during the pulse,
which allows details of the energy-flux spectrum to be followed in
time and thus characterize the spectral evolution. Such studies
have been made for instance by \citet{preece00} and \citet{RS02}.
However by necessity, the time-resolution becomes low and only the
strongest and longest pulses can be studied in this way due to the
need for high signal-to-noise ratio, SNR (\cite{preece98}). Using
this method, the spectral evolution has been quantified in
correlations between the observable parameters. The complete
spectral and temporal evolution of a pulse can be characterized by
three main observables, the peak of the $E F_{\rm E}$  spectrum,
$E_{\rm p}(t)$, the instantaneous energy flux, $F(t)$, and the
derived quantity, the energy fluence ${\cal E} (t) = \int ^t F
dt'$. The relations between these three observables are given by
two empirical correlations:

(i) the hardness-intensity correlation (HIC; see \cite{BR01}),
which relates the flux of the source to the 'hardness' of the
spectrum (here represented by $E=\Epk$). For the decay phase of a
pulse the most common behavior of the HIC is
\begin{equation}
F=\F0 (\Epk/\E00)^{\eta}, \label{eq:HIC}
\end{equation}
where $\E00$ and $\F0$ are the initial values of the peak energy
and the energy flux at the beginning of the decay phase in each
pulse and $\eta$ is the power law index. \cite{BR01} studied a
sample of 82 GRB pulse decays and found them to be consistent with
a power law HIC in at least 57\% of the cases and the power law
index, $\eta$ was found to have a broad distribution peaking
approximately at 2.0 : $\eta = 2.0 \pm 0.7$. It should be noted
here that even though the fits in some cases are not necessarily
statistically consistent with a power law, in principle all cases
do have a hard-to-soft trend that can be approximated by a power
law. Indeed, by simply lowering the required level of significance
of the power-law fit, \citet{BR01} showed that even 75 \% of the
pulses were consistent with a power law. \citet{RS02} did derive
alternative analytical descriptions of the HIC for pulses that do
not follow exactly a power-law. Referring to their Fig. 1, it is
however obvious that the power-law description can indeed be used
as an approximate description. Below, we will therefore use the
analytical expression, given by Eq. \ref{eq:HIC}, to describe the
general behavior of all pulses. This description will then be
exact for 60-70 \% of pulses.

(ii) the hardness-fluence correlation, $E_{\rm pk}({\cal E})$
(HFC; \cite{LK96}) which relates the hardness to the time running
integral of the flux, that is, the fluence. Equivalently, the HFC
represents the observation that the rate of change in the hardness
is proportional to the luminosity of the radiating medium (or,
equivalently, to the energy density);
\begin{equation}
\dot{\Epk} = -\frac{F}{\Phi_0}. \label{eq:HFC1}
\end{equation}
This corresponds to a linear relation between the $\Epk$ and the
time integrated energy flux, the energy fluence,
\begin{equation}
\Epk (t) = \E00 - \frac{{\cal E}(t)}{\P0} \label{eq:HFC}
\end{equation}
\noindent where $\P0$ is the decay constant. This behavior is most
often seen over the entire pulse and occurs in a vast majority of
pulses \citep{LK96, RS02}.

Ryde \& Svensson (2000)  showed that these two correlations (Eqs.
[\ref{eq:HIC}, \ref{eq:HFC}]) form a complete description of the
spectral evolution and define the light curve as well. In the
original description they showed that the observed photon flux,
$F(E)/E$ or $F(\nu)/\nu$, can often be described by a reciprocal
function in time. In \S \ref{sec:analytical} below we will
reformulate their results in terms of the energy flux.

\subsection{Low Resolution Spectroscopy (LRS)}

To utilize the full temporal resolution of, for instance, the
BATSE data, the deconvolution is not made and the incoming
light-curves are approximated by the count light-curves.  This is
also the case for weaker pulses for which the analysis described
in the previous section is not possible. The spectral evolution is
then instead quantified through the change in shape of the pulse,
in two or more broad energy bands. For instance, the four channels
of the BATSE discriminator rates (64 ms time resolution) can be
used, and {\it Beppo-SAX}, {\it Hete-II}/FREGATE, and {\it Swift}
have similar capabilities. Evolution of the spectrum, such as
variations in $\Epk$ and of the power-law slopes, will cause the
light curves in the different energy-channels to be different.

To quantify this change one can measure the shift in time between
light curve pulses (spectral lags) and their change in width.
These can be measured directly by using an analytical prescription
of the pulses, but more often they are measured with the use of
correlation functions (e.g., \citet{fen95}).
The {\it spectral} lag, $\Delta t$, is defined as the lag, $s$, at
which the cross correlation function (CFF) between the light
curves in two different {\it spectral} channels, $g$ and $h$, has
its maximum:
\begin{equation}
\rm{CCF} (g,h)(s)=\int_{- \infty} ^{\infty} g(\tau+s) h(\tau) \,
d\tau. \label{lag}
\end{equation}
\noindent If the two light curves are very similar to each other,
then the maximum occurs at a lag that measures the time that one
is shifted compared to the other.  For a detailed discussion on
the general characteristics of CCF, we refer to \citet{band97} who
gives the analytical expression of the CCF for a pure exponential
decay  (see his Fig. 1). Another way of quantifying the averaged,
spectral evolution of a burst, using the LRS data, is to measure
the hardness-ratio (HR) of the fluences (the time-integrated
fluxes) in two channels. The HR and the CCF are not independent
and their relation is often sought.

\citet{norris2} (see also \citet{norris3}) demonstrated that there
is a trend that the HR is anti-correlated with spectral lag. A
similar trend also exists for the peak flux, in that the long-lag
bursts all have low peak fluxes. Furthermore, the distribution of
lags is dominated by short lags, as also found by \citet{band97}.
A first step to understanding the connection between the detailed,
spectral evolution and the spectral lags was taken by
\citet{KL03}. They found that the HFC parameter $\P0$ has a nearly
linear correlation with lag. An even tighter correlation was found
when $\P0$ was normalized to the peak flux. Finally, the HR and
the $E_{\rm p}$ are naturally correlated. Indeed, \citet{BNB04}
argued for an empirical relation where $E_{\rm p}$ is proportional
to the square of the HR.

\citet{WF00} pointed out that it is not always reliable to
determine lags with CCF. This is especially the case for
multi-peaked bursts, for which both the HR and the CCF give an
average quantitative description. Variations of a burst's spectrum
occur both {\it within} a pulse as well as {\it between} pulses.
This was conclusively shown by \citet{band97}, who used
correlation functions and was able to conclude that $\sim 90 \%$
of pulses have a hard-to-soft trend, while $\sim 80 \%$ of the
multi-peaked bursts have a softening trend between peaks.
Varying $\E00$, $\eta$, and $\alpha$ between individual pulses
will be important in determining the integrated spectrum, which is
the spectrum that is quantified by the lags and the HRs.
\citet{WF00} also noted that the methods used to calculate the
CCFs can affect the results considerably. For instance, the
inclusion of time intervals when the signal is at background will
clearly affect the measurements. This adds complications for the
interpretation of these quantities. Below, we concentrate our
study  to an analytical model describing a single emission
episode, resulting in an individual pulse. A comparison of the CCF
measurements and the actual lags is further investigated in paper
II.

\section{Analytical Model}
\label{sec:analytical}

To catch the essence of the spectral evolution, we use the
following simplified model: the emission episode consists of a
single pulse which is assumed to be dominated by the decay phase,
called a FRED pulse (fast-rise-exponential-decay). We also assume
that the pulse has a monotonic decay of the spectral hardness, as
measured by $\Epk$, that is, it is a hard-to-soft pulse
(\cite{ford95}). The motivation to study such a model is, first,
that it is the individual pulses that bear the important physical
signatures of the emission processes and these signatures are
revealed mainly in the decay phase (\cite{RP02}). Most observed
pulses do indeed have such a shape, as noted by \cite{KRL} and
\cite{RBL03}. Second, most detailed spectral evolution studies
have been made on the decay phase, since it often constitutes a
major part of the pulse. Third, apart from the fact that many
pulses are observed to be hard-to-soft pulses, pulses for which
$\Epk$ is observed to track the flux could very well still be
intrinsically hard-to-soft. A motivation for this was given by
\cite{KRL}; pulses that are emitted prior to the analyzed one
often overlap it to some extent. Since, in general, the spectra of
pulses soften with time, this contamination will contribute mainly
soft photons. This will cause the measured spectrum to have an
$\Epk$ that is lower than that of the spectrum of the analyzed
pulse itself. Below, we derive the behavior of the decay phase of
a pulse and make a few analytical estimates of the dependence of
the lag on various variables.

\subsection{Decay Phase of a Pulse} \label{sec:decay}

An analytical expression for the  shape of a pulse can be derived
based on the knowledge of the spectral evolution. We follow the
calculations first outlined in \citet{RS00} (see also \citet{KRL})
to derive the energy-flux decay of a pulse. The combination of the
two empirical relations in Eqs. (\ref{eq:HIC}) and (\ref{eq:HFC})
gives the following differential equation governing the spectral
evolution
\begin{equation}
\dot{\Epk}=-\frac{F_0}{\phi_0 E_0^{ \eta}} \Epk^ \eta ,
\end{equation}
\noindent where $\dot{\Epk}$  is the time derivative of the peak
energy, $\F0$ and $\E00$ are the values of the flux and the peak
energy at the start of the decay phase. The solution gives the
energy flux
\begin{equation}
F(t) = \F0 \left(1+\frac{(\eta-1)t}{T} \right)^{-\eta/(\eta-1)}
\label{eq:Fdecay}
\end{equation}
\noindent and the corresponding peak energy
\begin{equation}
\Epk(t) = \E00 \left(1+\frac{( \eta-1)t}{T} \right) ^{-1/(
\eta-1)} \label{eq:Et}
\end{equation}
\noindent where we have defined
\begin{equation}
T \equiv \P0 \E00/\F0. \label{eq:Tdef}
\end{equation}
To analyze the behavior of this function in the case $\eta =1$, we
note that the denominators above both have the appearance of
\begin{equation}
\left( 1-\frac{t/T}{n}\right)^n , {\rm{with}} \;\; n \equiv
1/(1-\eta),
\end{equation}
\noindent This expression has a limiting value of $\rm{exp}(-t/T)$
as $n \equiv (\eta -1)^{-1} \rightarrow \infty$. Therefore as
$\eta$ tends to $1$, Eqs. (\ref{eq:Fdecay}) and (\ref{eq:Et})
become
\begin{equation}
F(t) =\F0 e^{-t/T}
\end{equation}
\begin{equation}
 \Epk(t) =  \E00 e^{-t/T}.
\end{equation}
This shows that the full behavior is described by Eqs.
(\ref{eq:Fdecay}) and (\ref{eq:Et}) {\it for all} $\eta$.  We also
note that it is possible to measure $\P0$ directly from the light
curve by identifying the time scale $T = \P0 \E00/\F0$, instead of
by the analysis of the spectra otherwise necessary.

Introducing $d \equiv \eta/(\eta-1) $ ($d$ as in the asymptotic
{\it d}ecay of the energy flux) we describe the peak and the
energy flux decays as
\begin{equation}
F(t) =   \F0 \left(1+\frac{t}{T(d-1)} \right)^{-d}\label{eq:7}
\end{equation}
\begin{equation}
\Epk(t)=   \E00 \left(1+\frac{t}{T(d-1)} \right)
^{1-d}.\label{eq:8}
\end{equation}
In \S \ref{sec:numsetup} we discuss an expansion of these
calculations to also include the rise phase, since we will need
the full description in the numerical study.

\subsection{Spectral Lags}\label{sec:analytical-lags}

Assume now that the two spectral channels between which the lag is
measured, are characterized by the energies $E_1$ and $E_2$.
Assume further that the dynamical range of the change in $\Epk$ is
larger or of the same order as $E_1/E_2$. These assumptions are
reasonable, which can be noted in \cite{RS02}, who studied the
decay of many pulse decays and found  that the typical dynamic
range was $ \sim 4 - 5$ (see their Fig. 1). Including a monotonic
energy decay during the rise phase of a pulse, this range will
increase. Furthermore, typical values for BATSE are $E_1= 25$ keV
and $E_2=100$ keV. With these assumptions a portion of the
non-power-law, curved photon spectrum will pass through both
channels. The pulse shapes (peak and width), detected in the two
channels, are then determined by how the spectral distribution
changes with time; the main change of the spectrum is in the decay
of $\Epk$. The spectrum will then pass through a certain channel
at a speed $d\Epk/dt$ which usually changes with energy. The
spectral lag can be thought of as the time it takes for the
spectrum to move from $E_1$ to $E_2$, which clearly is determined
by $d\Epk/dt$. Note that the actual peak, $\Epk$, does not
necessarily need to pass through the channels.

Before discussing the analytical consequences of this, we will
study three conceptually simple situations, which are valuable in
illustrating the discussion. First, consider a situation where
there is no spectral evolution, that is, the spectrum remains the
same during the change in intensity\footnote{This situation never
occurs for observed GRB pulses}; $F\propto \Ep ^{\eta}$ with
$\eta=\infty$. Then there will be no difference in the light curve
between the bands, apart from a normalization, and the spectral
lag would be zero. Second, consider a situation
 where we allow for a spectral evolution, but approximate
the spectra with a Dirac $\delta$-function in energy (see, e.g.,
\cite{cen99} and \cite{RS99}) so that $F_{\rm E}(E,t) = F(t) \cdot
\delta [E-\Epk(t)]$, that is, with a monochromatic spectrum. In
this situation, it is clear that the time interval between the
pulse peaks in the two channels is exactly the time that the
spectrum takes to move from $E_1$ to $E_2$. A third situation,
that is still more realistic,  is to take an energy spectrum that
is described by a Heavyside function, $F_{\rm E}(E,t) = F(t) \cdot
\Theta [E-\Epk(t)]$, that is, the spectrum consists of  two power
laws with photon indices $\alpha = -1$ and $\beta= -\infty$ that
are sharply joined at $\Epk$. As the break of the spectrum passes
a certain energy, the flux disappears, which can constitute a
characteristic point in the light curve. The difference in channel
light-curves again will clearly reflect the $\Epk$ evolution.

Based on these discussions, we therefore make the following
simplified model that captures the general behavior, namely that
the lag, $\lag$, is determined by
\begin{equation}
 \Dt = \int_{\Delta E} \frac{dt}{d\Epk} \,dE
 \label{eq:15}
\end{equation}
and the rate, $d\Epk/dt$, is determined by the spectral evolution
parameters according to Eq.(\ref{eq:Et}). Note also that,
according to Eq. (\ref{eq:HFC1}) this rate is proportional to
$F/\P0$ (see also \citet{KL03}). To describe this analytically, we
assume that the above analytical description of the evolution of
$\Epk$ is also valid for the rise phase of the pulse. Eq.
(\ref{eq:Et}) then  gives
\begin{equation}
\frac{t}{T}= \frac{1}{\eta-1}\left(\left[
\frac{\Enoll}{E}\right]^{\e1} -1 \right),
\label{eq:a1}
\end{equation}
Here, $\E00$ represents the maximal value of $\Epk$, that is, the
value at the start of the rise phase in contrast to the definition
in \S \ref{sec:decay}. This equation is also valid for all $\eta$,
since as $\eta$ tends to $1$, Eq. (\ref{eq:a1}) has a limiting
value of
\begin{equation}
\frac{t}{T} =  \ln \frac{\Enoll}{E}.
\end{equation}
If the spectral evolution of a pulse is governed by this law, then
the spectral lag, $\Dt$, between two energy channels with
characteristic energies $E_1$ and $E_2$, will be given by
\begin{equation}
\frac{\Dt}{T} = \left(\frac{\Enoll}{E_2}\right)^{\e1}
\frac{1}{\e1}\left(\left[\frac{E_2}{E_1}\right]^{\e1}-1\right).
\label{eq:18}
\end{equation}
Here $E_1 < E_2$ and a positive lag is defined as the hard
radiation precedes the soft. The ratio ($E_2/E_1$) in this
equation is an instrumental constant and we denote the last two
factors  in Eq. (\ref{eq:18}) by $\psi (\e1)$, which tends to
$\psi(0) = \ln(E_2/E_1)$ as $\eta$ tends to 1. For a specific
instrument the lag can thus be  written as
\begin{equation}
{\Dt} = T \cdot (\Enoll/E_2)^{\e1}  \cdot \psi(\e1).
 \label{eq:19}
\end{equation}
\noindent Apart from the linear dependence on $T$, the lag depends
on the value of $\Enoll$ relative to the high energy channel, as
well as to the instrumental function $\psi(\e1)$. In Eq.
(\ref{eq:19}) one could have chosen to include the second factor
into the instrumental function. However, the energy band of a
certain instrument limits the observed values of $E_0$, which
therefore will be correlated with the typical energy of the band.
Therefore, it is useful to keep this factor separate in such a
comparison. Finally, using the definition of $T$, in Eq.
(\ref{eq:Tdef}), we find that
\begin{equation}
\Dt = \P0 \frac{\Enoll^{\eta}}{F_0} \cdot \frac{\psi
(\e1)}{E_2^{\e1}}, \label{eq:linear2}
\end{equation}
 which, as $\eta$ tends to 1, has a limiting value
\begin{equation}
{\Dt} = \P0 \frac{\Enoll}{F_0} \cdot \psi \left( 0 \right).
\end{equation}

  \begin{figure}
   \centering
  \includegraphics[width=7cm]{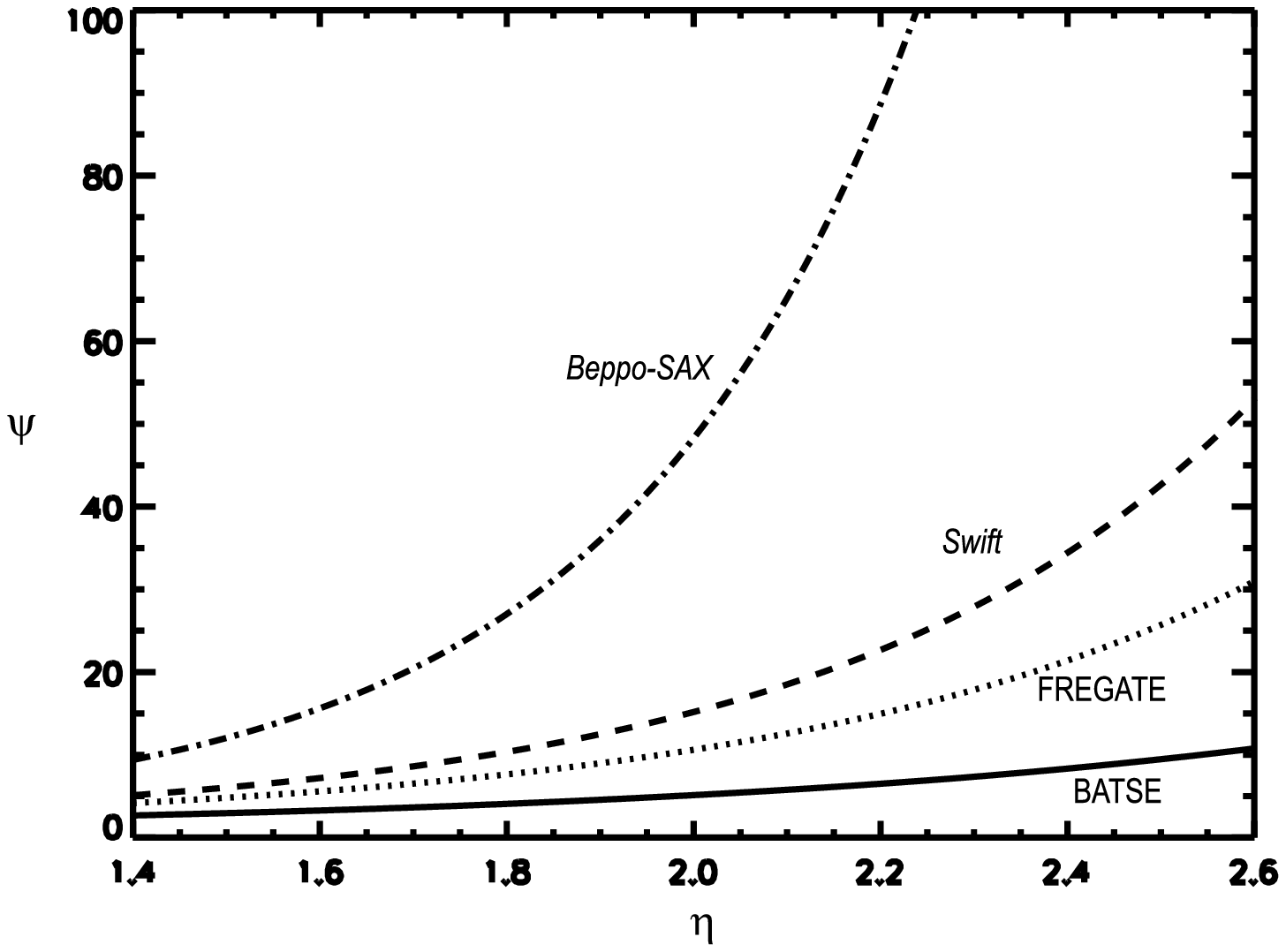}
  \caption{Instrumental-dependent factor, $\Psi$, as a function
 of the HIC index $\eta$, according to Eq. [\ref{eq:19}].
 The solid line is for BATSE, the
 dotted line is for FREGATE, the dashed line is for {\it Swift},
 and the dot-dashed line is for {\it Beppo-SAX}}
              \label{fig:psi}
   \end{figure}

In Fig. \ref{fig:psi}, the function $\psi$ is plotted versus the
HIC power-law index $\eta$, for four different detectors, BATSE,
{\it Beppo-SAX}, FREGATE and {\it Swift}. For BATSE, $\psi$ varies
only weakly while for the other three instruments it varies
somewhat more. For instance, as $\eta$ changes from  1.5 to 2.5,
$\psi$ varies by a factor of 3.2 (BATSE), 5.3 (FREGATE), 7.1 ({\it
Swift}), and 19 ({\it Beppo-SAX}), respectively. The typical
energies used were $E_1/E_2$ = 32.5/200 (BATSE), 18.5/215
(FREGATE),  5.1/82.5 ({\it Swift}), and 7.5/370 ({\it Beppo-SAX}).
The variation of $\psi$ for BATSE is relatively small, while
spectral lags vary over several orders of magnitude
\citep{norris2}. Furthermore, Ryde \& Svensson (2002) showed that
$E_0$ cluster narrowly in $<E_0>= 332 \pm 45$ keV for BATSE
bursts. The conclusion is, therefore, that the lag is determined
mainly by the time-scale of the light curve. However, this
relation is probably weaker for {\it Beppo-SAX}, FREGATE and {\it
Swift} bursts, since they have a stronger dependence on $\eta$,
which will cause a larger dispersion. This is the case in
particular for the large energy interval used in {\it Beppo-SAX}.

An alternative, compact derivation of the above, containing the
same information is the following:
\begin{eqnarray}
\Dt & = & \int \frac{1}{\dot{E}} \,\, dE = - \int \frac{\P0}{F} \,
dE
 =  \P0 \int (kE^{\eta})^{-1} \,\, dE \nonumber \\
& = & \frac{\P0}{k\,(1-\eta)} \left( E_2^{1-\eta} -E_1^{\eta-1}
\right)
\end{eqnarray}
\noindent where we first used the HFC (Eq. [\ref{eq:HFC1}]) and
then used the HIC (Eq. [\ref{eq:HIC}]).

\subsection{Improvements of the Analytical Model}

In the discussion above, the change in the channel light-curves is
assumed to be captured by the $\Epk$ decay, and the change in flux
is assumed not to alter the lag significantly. The full
description is, however, given by
\begin{eqnarray}
F_{\rm ch}(t) & = & F_{\rm bol} (t) \int_{\rm ch} {\cal{B}} (E,t)
\, dE \nonumber \\
 & = & F_{\rm bol} (t) \Epk (t) \int_{x_{\rm min}} ^{x_{\rm max}}
{\cal{B}} (x) \, dx \label{eq:full}
\end{eqnarray}
Here, ${\cal{B}} (E,t)$ describes the instantaneous spectrum, for
instance, by the empirical GRB function (\cite{band}) and $F_{\rm
bol} (t)$ is the bolometric energy flux, that is, the flux
integrated over the whole spectrum. As mentioned above, we assume
a monotonically decaying $\Epk$ and in the last step in the above
equation we note that the energy dependence of ${\cal{B}}(E,
t(\Epk))$ is normally in the ratio $x \equiv E/\Epk$.
Note that $\int_0 ^\infty {\cal{B}} \, dx = 1/\Epk$. The full
spectral evolution is easily treated numerically by using Eq.
(\ref{eq:full}), and this will be done in \S
\ref{sec:simulations}. These results are later compared with the
above, simplified analytical model.

There are two further issues that could be taken into account.
First, in the above analytical treatment the low-energy spectral
slope is assumed to be constant. This is indeed the case in many
pulses, see for instance Figs. 3 and 5 in \cite{RS99}. However, in
other cases the slope can change significantly, often with a
softening trend. This can be seen Fig. 2 in \cite{crider97}.
Second, the pulse could actually be a tracking pulse which would
need a more elaborate analytical approach. Finally, the $\Epk(t)$
behaviors used in the above derivations are the most commonly
observed. However, it could be useful to study other behaviors.
However, all these issues will lead mainly to minor corrections of
the main behavior and they will be addressed elsewhere.

In summary, the first conclusion that can be drawn from the
analytical model is that the time lag measures a combination of
all the parameters describing the spectral evolution, $\P0, \eta,
E_0$ and $F_0$. Specifically, Eq.(\ref{eq:19}) shows that the lag
has a linear relation to the pulse decay time scale, $T$. In an
alternative formulation there is also a linear dependence on
$\P0$, as in Eq.(\ref{eq:linear2}). \citet{KL03p,KL03} found that
the lag is correlated to the $\Phi_0$, see also discussion in
\citet{schaefer04}. A stronger linear correlation should emerge
between $\Dt$ and $\P0/\F0$. If $\psi(\e1)$ and the ratio
$\E00/E_2$ varies only slightly, such a correlation will emerge.

\section{Numerical Simulation}
\label{sec:simulations}

The above analytical treatment of the spectral evolution is
useful, even though a few approximations had to be made. Here, we
will numerically simulate burst pulses using the full description
in Eq. (\ref{eq:full}) and measure the spectral lags. We will use
the knowledge gathered from the high-resolution spectroscopical
investigations, described in \S \ref{sec:HRS}.

\subsection{Numerical Setup}
\label{sec:numsetup}

 \begin{figure*}
\centering
\includegraphics[width=\textwidth]{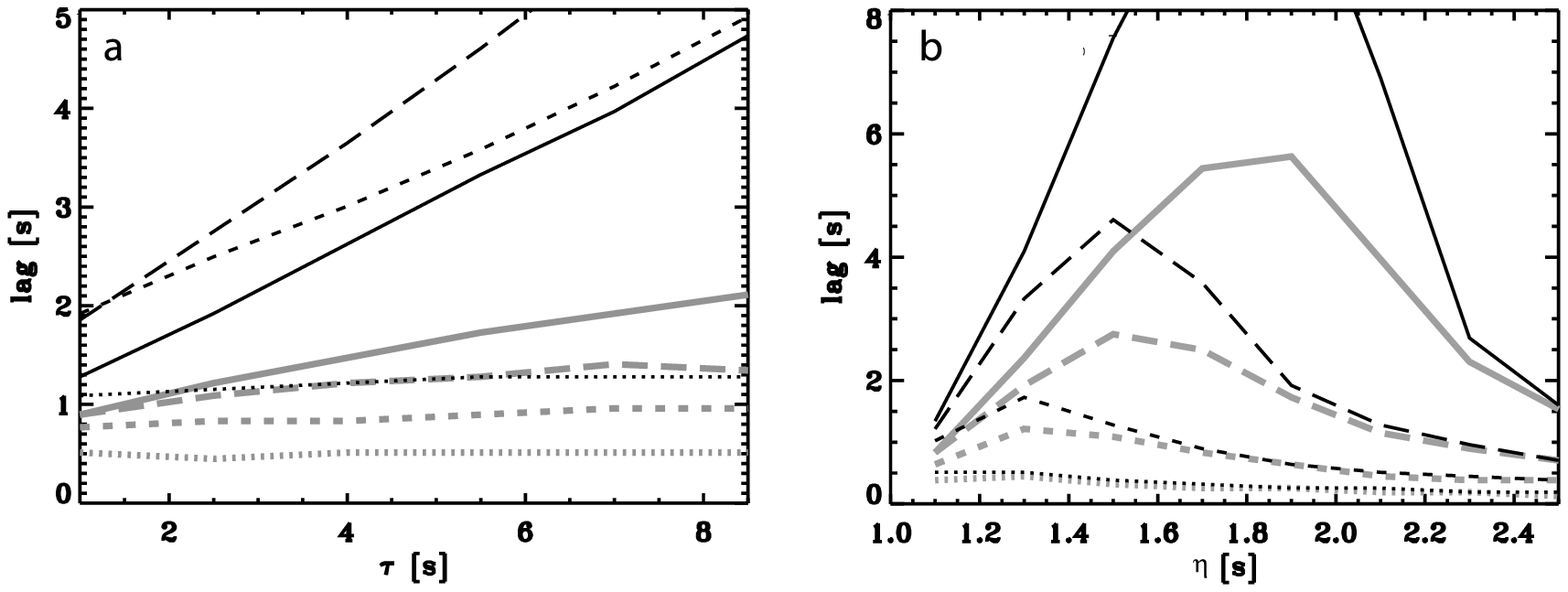}
 \caption{(a) Spectral lag as a function of pulse decay time
scale, $\tau$, showing near-to linear relations. The parameters
determining the spectral evolution were $\E00 =500$ keV, $\tr=1$
s, $\alpha=0$ (black lines) and $\alpha = -0.5$ (grey lines). The
HIC power-law index ($d\,{\rm log}F/d\,{\rm log}\Epk$) $\eta= 1.3$
(solid), $\eta= 1.5$ (long dashed), $\eta= 1.7$ (short dashed),
$\eta= 2.1$ (dotted). (b) Spectral lag as a function of $\eta$.
Here $\E00 = 500$ keV, $\tr=1$ s, $\tau=5.5$ s (black) and $\tau =
2.5$ s (grey). The low-energy power-law index, $\alpha= 0.5$
(solid), $\alpha=0.0$ (long dashed), $\alpha = -0.5$ (short
dashed) and $\alpha=-1.0$ (dotted). \label{fig:tau_and_eta}}
 \end{figure*}

 \begin{figure*}
\centering
 \includegraphics[width=\textwidth]{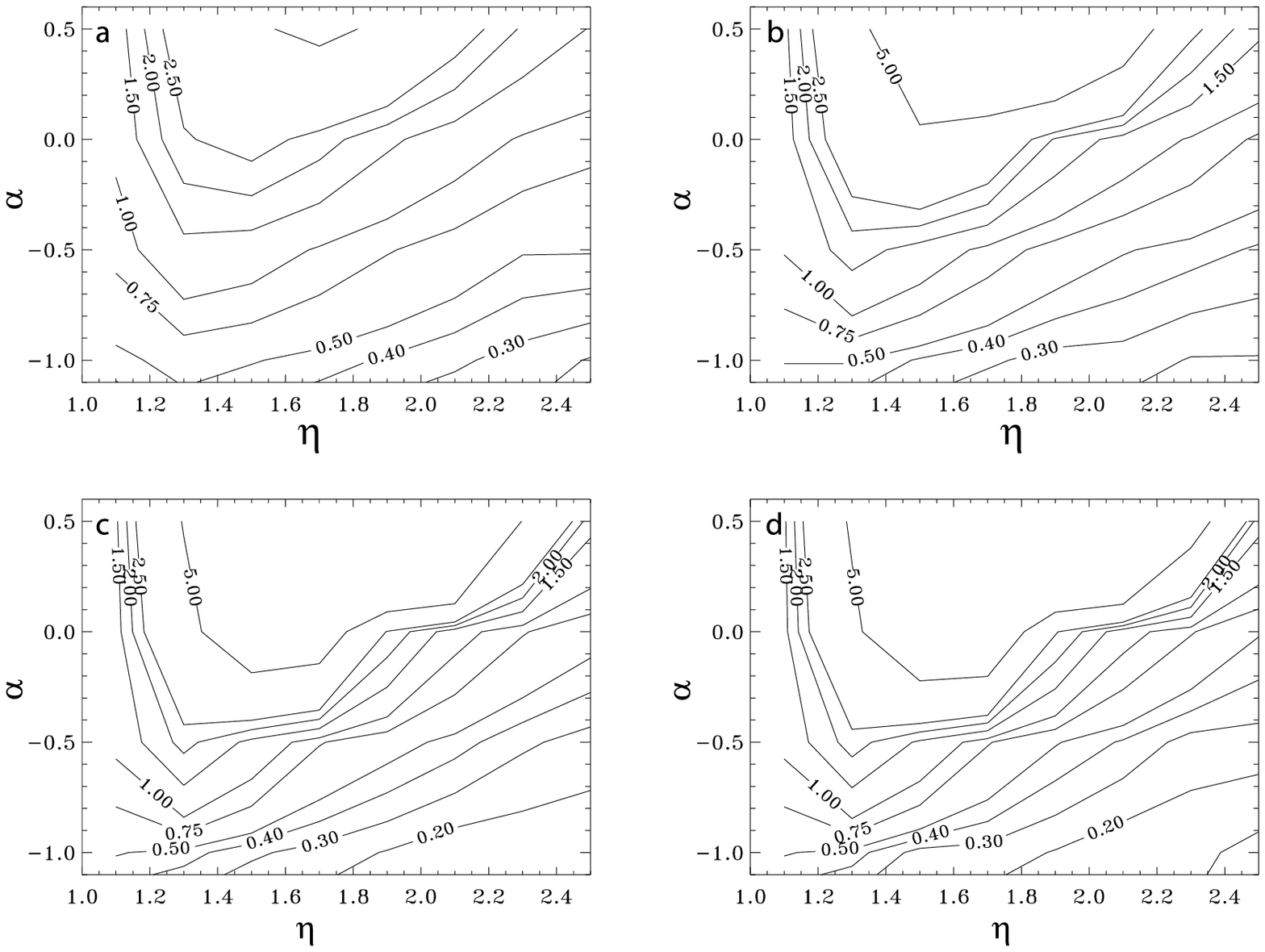}
 \caption{Contour plots of the spectral lag as a function of
 the HIC power-law index, $\eta$ and the spectral, low-energy,
 power-law, $\alpha$ for varying values of $\E00$: (a)  $\E00 =
 100$ keV, (b)  $\E00 = 250$ keV, (c)  $\E00 = 500$ keV, (d)  $\E00 =
 750$ keV.
 \label{fig:contour_eta_alpha}}
 \end{figure*}

  \begin{figure*}
\centering
 \includegraphics[width=\textwidth]{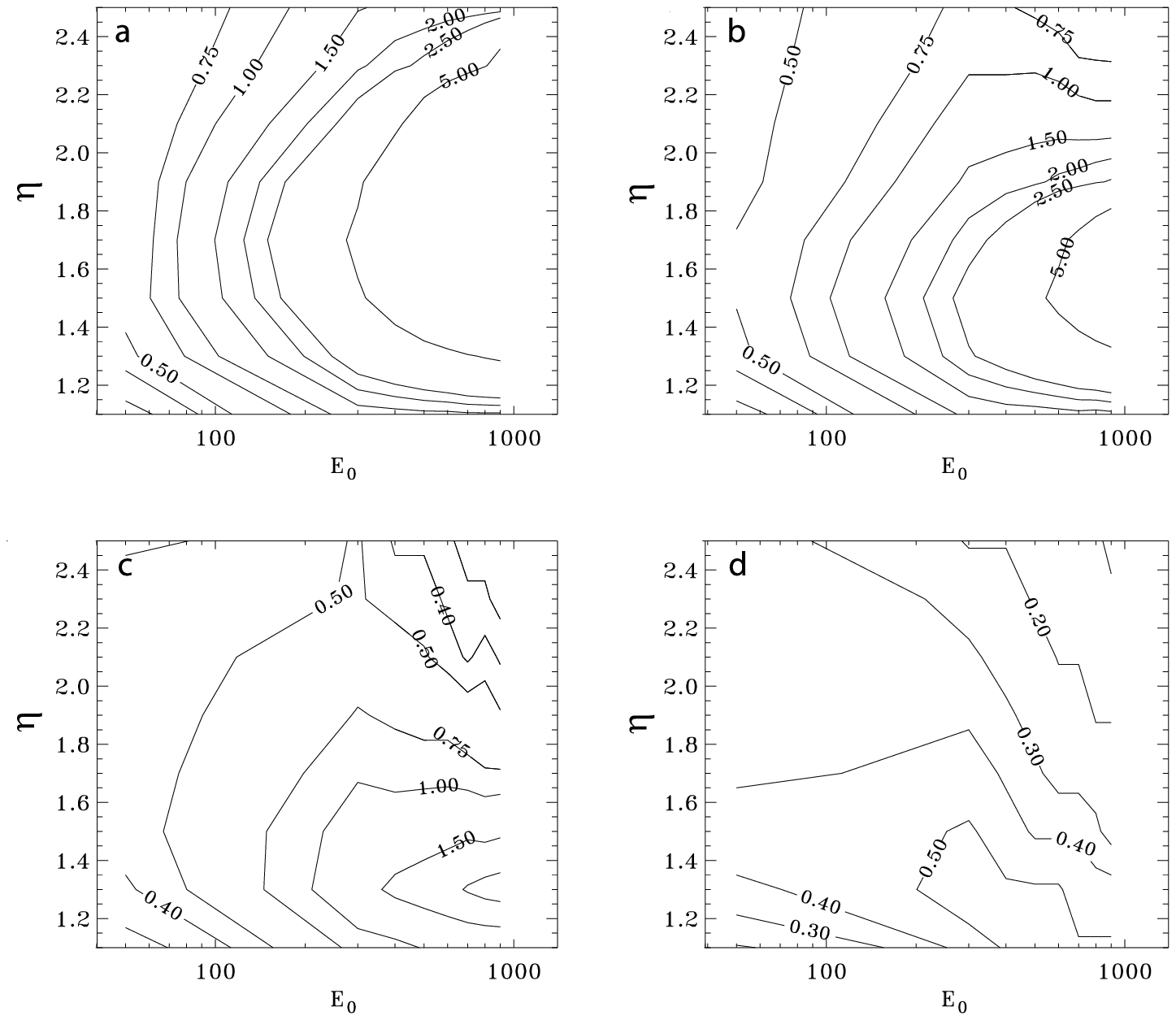}
 \caption{Contour plots of the spectral lag as a function of
 the HIC power-law index, $\eta$ and the energy $E_0$,
  for varying values of $\alpha$: (a)  $\alpha =
 0.5$, (b)  $\alpha = 0.0$, (c)  $\alpha = -0.5$, (d)  $\alpha =
 -1.0$. These plots correspond to the black lines in Fig. 2b.
 \label{fig:contour_eta_E}}
 \end{figure*}

Kocevski, Ryde, \& Liang (2003) expanded the analytical study  of
the pulse shape, made above, to also include the rise phase. We
review this treatment here as we will make use of their result in
the numerical simulations. As the evolution of $\Epk$ is still
from hard to soft during the main part of the {\it rise phase} of
a pulse, the energy flux and the $\Epk$ are anti-correlated during
this phase. By arguing from physical first principles \cite{KRL}
studied several analytical shapes for the whole pulse that
includes the rise phase and asymptotically approaches
Eq.~(\ref{eq:7}) in the decay phase (which is the behavior we are
interested in here). In most physical models both the peak of the
energy spectrum and the luminosity are proportional to the random
Lorentz factor of the shocked electrons to some power: $\Epk (t)
\propto \Gamma_{\rm r}^a (t) \cdot g(t)$, and $ F_{\rm E}(t)
\propto \Gamma_{\rm r}^b (t) \cdot h(t). $ The functions $g(t)$
and $h(t)$ parameterize the unknown time dependence on particle
densities, optical depth, magnetic field, kinematics, etc. The
correlation between hardness and the energy flux can thus be
described as
\begin{equation}
F_{\rm E}(t) \propto \Epk ^{\gamma} \cdot f(t)
\end{equation}
\noindent with $\gamma = b/a$ and $f(t)=h(t) g(t)^{-\gamma}$.
During the decay phase of a pulse the power-law relation dominates
the HIC, as in Eq.[\ref{eq:HIC}], whereas during the rise phase it
will be dominated by the unknown function $f(t)$. It is reasonable
to assume that the rise phase is connected to some transient
process, for instance, an initial decrease in optical depth,
increase in the number of energized particles, or the merging (or
crossing) of two shells. After this initial phase, the original
decay behavior as described by Eqs. (\ref{eq:7}) and (\ref{eq:8})
should emerge. We therefore use the following prescription

\begin{equation}
f(t) \propto 1-e^{-t/\tr}
\end{equation}

\noindent where the time constant $\tr$ now represents the {\it
r}ise phase. This corresponds to $h(t) \propto g(t) ^{\gamma}$ as
$t >> \tr$ and the function reaches the power law in
Eq.(\ref{eq:HIC}) asymptotically. A few new parameters have to be
introduced; the energy flux, the peak energy, and the time of the
peak of the light curve, $\Fm$, $\Em$, and $\tm$:
\begin{equation}
F=\Fm  \left( \frac{\Epk}{\Em} \right)^{\gamma}
\frac{1-e^{-t/\tr}}{1-e^{-\tm/\tr}} \label{equ:F2}
\end{equation}
which, combined with the HFC (Eq.[\ref{eq:HFC}]), gives the
differential equation governing the spectral and temporal
evolution of the pulse. It has the following solution (see
\cite{KRL} for details).
\begin{equation}
F(t)=\frac{A_0 (1-e^{-t/\tr})}{(1+(t+\tr e^{-t/\tr})/\tau)^d}
\label{F2}
\end{equation}
\begin{equation}
\Epk(t)=\frac{A_1}{(1+(t+\tr e^{-t/\tr})/\tau)^{(d-1)}}
\end{equation}
where $A_0$ and $A_1$ are analytical functions of $d, \tau, \tr,
\P0, \Fm, \Em$ and $\tm$ and the decay index $d$ is defined by
requiring that $F(t) \rightarrow t^{-d}$ as $t \rightarrow
\infty$, which gives $\eta=d/(d-1)$. We introduce the peak energy
at $t=0$, $\E00$ which gives $A_1= \E00
(1+\tr/\tau)^{(1/\eta-1)}$. The value of the flux at $t=0$ is
arbitrary and is not important. The parameter $d$ is found to be
about 2-3 \citep{KRL}.

The instantaneous {\it photon spectrum} is modelled by the
standard GRB model (\cite{band}), which is essentially a
low-energy power-law, $F\propto E^{\alpha}$ exponentially
connected to  a high-energy power-law $F\propto E^{\beta}$ where
the photon indices $\beta  < \alpha$. In an $E F_ {\rm E}$
representation the spectrum peaks at $\Epk$. The value of $\beta$
was kept constant throughout the analysis here, since we
concentrate our efforts to study the dependence of the lag on
$\alpha$.

To be able to study spectral lags, the light curves for the four
BATSE channels were found by integrating the spectra over the
respective bands.  Normally, the energy edges vary somewhat from
observation to observation. For the simulation we used typical,
average values: $E_1= 24$ keV, $E_2=60$ keV, $E_3=108$ keV,
$E_4=10000$ keV. The spectral lag was then found by measuring the
time between the peaks of the light curves in channel 3 and
channel 1. The parameters that are used as input to the numerical
analysis are the following: the maximal peak energy $E_0$, $F_0$,
the low- and high-energy power laws, $\alpha$ and $\beta$, the HIC
index $\eta$, the decay- and rise- time scales, $\tau$ and $\tr$.

\subsection{Spectral lags}

The solid lines in Fig. \ref{fig:tau_and_eta}a show the spectral
lag, $\lag$, as a function of the decay time scale $\tau$ with
$\alpha =0$, and for four different values of $\eta=1.3, 1.5,
1.7,$ and $2.1$. The dashed lines show the corresponding behavior
for $\alpha=-0.5$. For both cases, $\E00= 500$ keV and $\tr= 1$ s.
The relation between $\lag$ and $\tau$ is approximately linear
(except for the case of $\eta =2.1$, where a maximal value is
reached). This shows that the lag is closely related to the decay
timescale of the pulse. This is in agreement with the analytical
results in the previous section.

The dependence of the spectral lag with the HIC power-law index,
$\eta$, is shown in Fig. \ref{fig:tau_and_eta}b, for four
different values of $\alpha = 0.5, 0.0, -0.5, -1.0$. The solid
curves are for $\tau = 5.5$ s and the dashed are for $\tau = 2.5$
s. Again $\E00 = 500$ keV and $\tr = 1$ s.  The lag has a maximal
value for a certain $\eta$-value, which increases for increasing
$\alpha$. This illustrates the complexity of the interpretation of
the spectral lag: a certain lag does not uniquely correspond to a
$\eta$ value. The longer decay time compared to the rise time does
not change the general behavior, it merely moves the curves to
larger lags and larger $\eta$-values.

A more general picture of the dependence is given in the two
following figures. In Fig. \ref{fig:contour_eta_alpha} the contour
plots of $\lag$ is shown as a function of the HIC index, $\eta$,
and $\alpha$. The largest values are found for hard spectra (large
$\alpha$) and averaged HIC slopes. An increase in $\E00$,
increases the lag values mainly for the hardest spectra. Fig.
\ref{fig:tau_and_eta}b represents slices of these contours. The
corresponding plots with a dependence on the maximal energy,
$\E00$ is shown in Fig. \ref{fig:contour_eta_E} for $\alpha=0.5,
0.0, -0.5$ and $-1.0$ (left to right). Both Figs.
\ref{fig:contour_eta_alpha} and \ref{fig:contour_eta_E} are for a
pulse with $\tr=1$ s and $\tau=5.5$. As $\alpha$ becomes softer
(decreases) the lag, for a certain $\eta$ and $\E00$, decreases.
Also the maximal value occurs for smaller and smaller $\eta$
values. We have made similar runs with varying pulse asymmetry,
that is, varying the ratio of the rise and decay time-scales. As
the decay time-scale, $\tau$, increases relative to $\tr$, the
contour plots have in general a similar form except near the
maximal value of the lag, that is, the peak of the contour plot.
The peak increases and becomes more accentuated while the
parameter space away from the peak is constant.

\subsection{Hardness Ratios}

\begin{figure*}
\centering
\includegraphics[width=\textwidth]{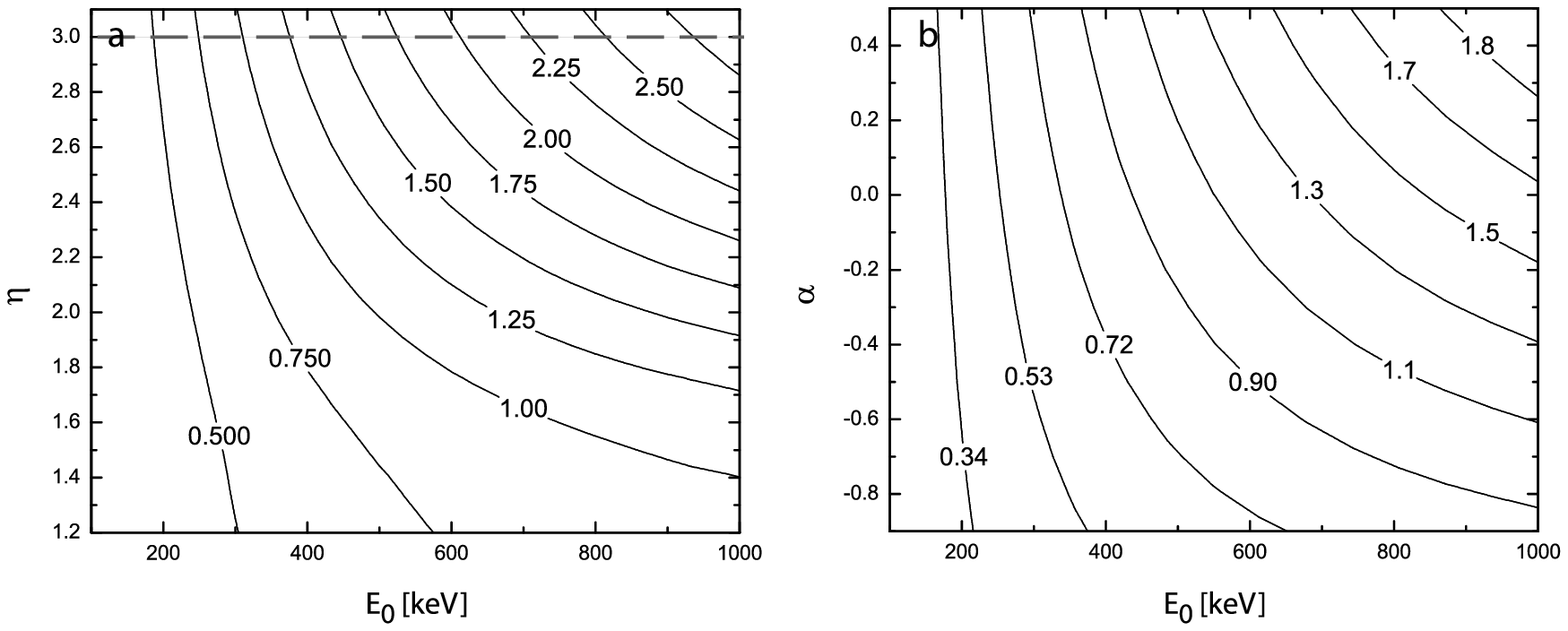}
\caption{Hardness ratio dependence on $\eta$, $\alpha$, and $E_0$,
shown as contour plots. In panel (a) $\alpha = 0$, while in (b)
$\eta = 2.0$. Note that the HR is practically independent of both
$\eta$ and $\alpha$ at low energies.
 \label{fig:HR}}
 \end{figure*}

We also studied the dependence of the hardness ratio (HR) on
varying $\E00$, $\eta$, and $\alpha$ over a single pulse. We
simulated pulses and calculated the integrated photon flux in
BATSE channels 1 and 3, over the interval of the pulse which was
brighter than 0.1 of the peak flux. The ratio of these fluxes is
defined as HR31. The results for a pulse with a rise time of 0.5 s
and decay time of 2 s are depicted in the following three figures.
Panel (a) in Fig. \ref{fig:HR} shows the contour plots of the HR
as a function of $\eta$ and $E_0$ while $\alpha$ was fixed and set
to 0.0. For low values of $E_0$ the HR13 is practically
independent of $\eta$ and is thus mainly determined by $E_0$. This
is the case in particular for large $\eta$. A corresponding
behavior is seen in panel (b) which shows the HR31 contours as a
function of $\alpha$ and $E_0$. Here $\eta$ was fixed instead and
set to 2.0. The hardness ratio HR13 is practically independent of
$\alpha$ for low values of $E_0$, most clearly pronounced for the
hardest spectra. The strongest dependence of the HR is therefore
on $\E00$.

\begin{figure*}
\centering
\includegraphics[width=\textwidth]{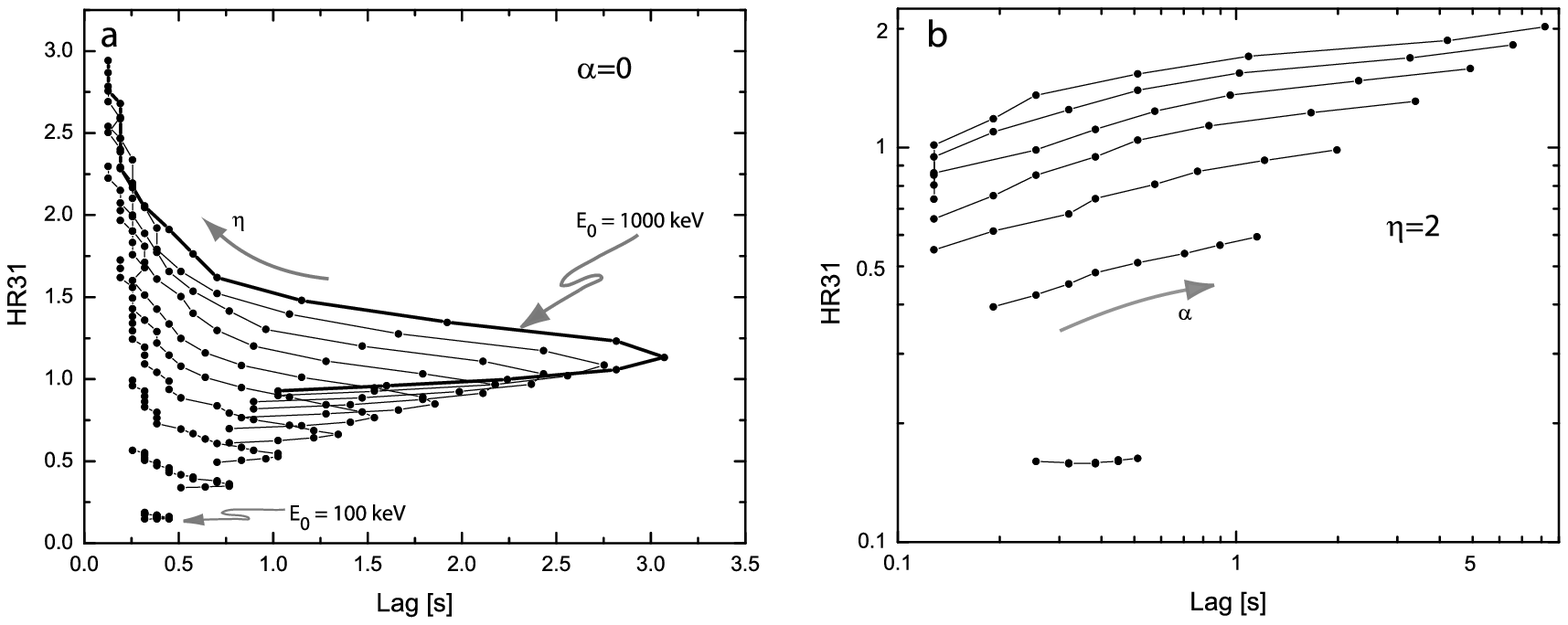}
\caption{Hardness ratio as a function of spectral lag.  In panel
(a) each curve is for a certain $\E00$, with the thick line
representing the largest value; $\eta$ increases upwards along the
lines. Similarly, in panel (b), $\E00$ increases upwards for each
curve and $\alpha$ increases to the right along each curve. In
panel (a) $\alpha = 0$, and (b) $\eta = 2.0$.
 \label{fig:HR2}}
 \end{figure*}

\begin{figure}
\centering
\includegraphics[width=0.5\textwidth]{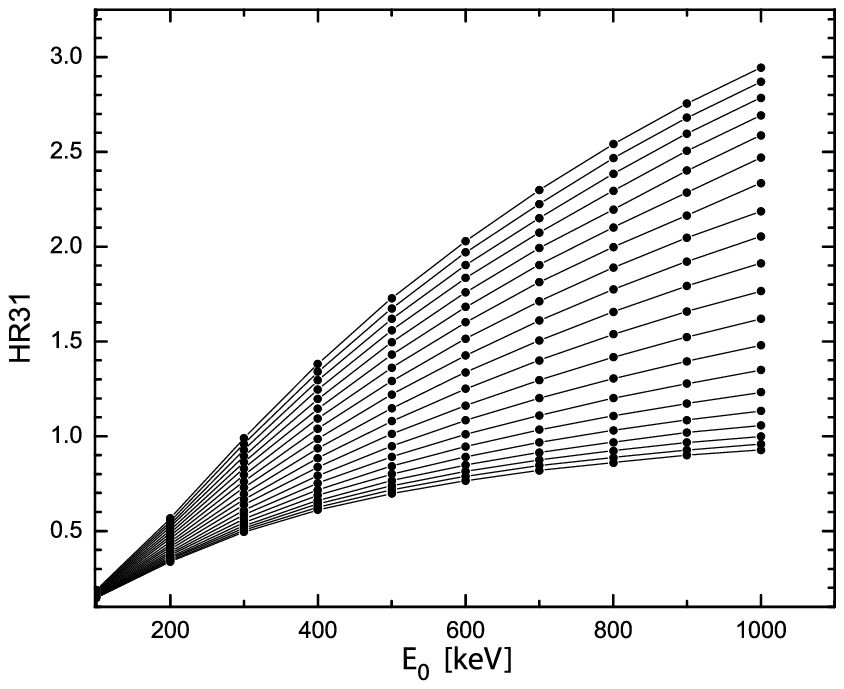}
\caption{The dependence of the hardness ratio $HR31$ on the peak
energy $E_0$, for the same run as in Fig. \ref{fig:HR}a. Each
curve is for a different $\eta$, increasing upwards in the figure,
from $\eta = 1.2$ to $3.1$. A variation in $\eta$ will thus lead
to a dispersion, which becomes larger for higher $\E00$-values.
 \label{fig:HR3}}
 \end{figure}

Often the relationship between the HR and the spectral lag is
investigated and such plots are often used in LRS work (see, e.g.,
Fig. 4 in \cite{norris2}). In panel (a) in Fig. \ref{fig:HR2} the
HR31 is plotted versus lag for a run with $\alpha$ set to 0.0. The
data points are connected for constant $E_0$, with $E_0 = 1000$
for the curve furthest to the right and $E_0= 100$ for the curve
furthest to the left. The index $\eta$ increases upwards in the
figure, along the curves, from $\eta = 1.2$ to $\eta = 3.1$. For
low $E_0$ the hardness ratio is seen to be practically constant,
independent of $\eta$, consistent with Fig. 5a. For large
$\eta$-values the dependence of HR on lag becomes greater (the
vertical section of the curves). The large change of the lag as
$\eta$ varies for a constant $\E00$ is also caught by Fig. 4b, as
discussed above. In panel (b) the HR31--spectral lag relations are
shown for varying $\alpha$ with $\eta$ set to 2.0. The relations
are approximate power-laws. This is consistent with previous
figures; in Figs. 3, for a constant $\eta=2$ and $E_0$, the lag
has a monotonic increase as a function of $\alpha$, while in Fig.
\ref{fig:HR}b, again with $\eta = 2$ and a constant $E_0$, we see
that the HRs as a function of $\alpha$ will have a monotonic
increase, or will be constant (independent). Hence the
correlations in panel Fig. \ref{fig:HR2}b. Here, a larger $\E00$
corresponds to a power-law at higher HR31 and the spectra get
harder (larger $\alpha$) to the right in the figure. Again, for
low $\E00$ values, the hardness ratio is independent of variations
in lag. The HR31 and $E_0$ are correlated which can be seen in
Fig. \ref{fig:HR}a and b. Among others, \cite{BNB04} have shown
that such a relation exists observationally between $E_{\rm p}$
and the hardness ratio. This is shown in more detail in Fig.
\ref{fig:HR3}, which is from the same run as Fig. \ref{fig:HR}a;
$\eta$ increases upwards, from $\eta = 1.2$ to $\eta= 3.1$. The
dispersion, introduced by a variation in $\eta$, is thus largest
for large values of $\E00$.

\section{Discussion}

The above numerical work has presented the somewhat complex
expression that the spectral evolution has in the two observables,
namely the spectral lag and the hardness ratio. The results can be
used to interpret lag and HR observations of various burst
samples.

\subsection{Interpretation}

The dependence of the lag on $\alpha$ is given in Fig. 3. It shows
that the lag increases with $\alpha$ for all tested values on
$\eta$. The more energetic the pulses/bursts are (higher $\E00$)
the steeper this correlation will be. Furthermore, the size of the
lags can give a hint of the hardness of the spectra: a hard
spectrum, for instance a thermal spectrum ($\alpha = + 1$) or a
small pitch-angle synchrotron-spectrum ($\alpha = 0$), will have a
larger lag than a corresponding softer spectrum, maybe produced by
optically-thin synchrotron emission ($\alpha = -2/3, -3/2$).
Figure 3 also shows that, if a sample of spectrally hard pulses
(large $\alpha$) is chosen, then the plot between lag and $\eta$
will have a characteristic bump indicating the distribution of
$\E00$ in the sample studied. Note that each panel in Fig. 3
corresponds to a specific $\E00$-value. A softer sample will have
a less pronounced dependence (compare Fig. 2b).

Furthermore, by studying the lag as a function of $\E00$, for a
sample of bursts, conclusions can be drawn using the information
available in Fig. 4. For instance, if the lag is found to increase
with $\E00$, the sample is probably dominated by hard pulses, or
at least do not have particularly large $\eta$-values. Oppositely,
the lag could be found to decrease with $\E00$. This could
indicate that the sample is dominated by soft spectra (panel d;
especially at high energies). For harder spectra, such a decline
is found only in pulses with the very largest $\eta$. Continuing
with Fig. 4, we see that if, once again, lag is plotted versus
$\eta$ and only a weak dependence is found, the pulses are most
probably low-energetic bursts, such as X-ray flashes. This is so
especially in the case of soft spectral pulses (compare Fig. 3).
For high-energetic bursts (high $\E00$) the characteristic
dependence, with a peak at some intermediate $\eta$, will be seen
and the range of observed lags will be large. For very soft pulses
(small $\alpha$) it will mainly be a declining function. Note that
the panels in the figure are for constant $\alpha$-values.
Correlations between, for instance, $\eta$ and $\alpha$ in the
sample under investigation will complicate the interpretation
slightly, since the same panel cannot be used for all $\eta$.
Also, since there are different possible dependence, the
distribution of $\E00$ of the sample will be important in
determining the final relation.

Similarly, differences in HIC-index $\eta$ between samples will be
marked by that the hardness ratio generally is largest for the
large--$\eta$ pulses (except for very low $\E00$). This is for
instance manifested in Fig. 7 in which the large--$\eta$ pulses
have a steeper dependence on the peak energy. The largest lags are
found for pulses with intermediate $\eta \sim 1.5$ -- $\eta \sim
2$. As mentioned above, soft burst will have a characteristic
switch in lag versus $\E00$ behavior, from increasing (for low
$\eta$) to decrease (for large $\eta$).

Figures 5, 6 and 7 reflect the behavior of the hardness ratios.
If, for instance, the HR is plotted versus $\alpha$ ($\eta$) and
the relations are found to be weak, it is again an indication of
low-energetic bursts. The distribution of the pulses in the
HR-lag--plane will give indications of both $\E00$ and $\eta$
according to the plot.

\subsection{Radiation processes and emission sites}

Several different emission mechanisms are probably active during
the burst (see e.g. Ryde 2004), most importantly various versions
of synchrotron (and/or inverse Compton) emission, predominantly
its optically-thin version, and thermal, black-body emission.
Other possible radiation mechanisms include synchrotron emission
from electrons with a small pitch-angle distribution (or jitter
radiation) and saturated Comptonization. The most visible
difference is in the hardness of the spectra, as measured by the
low-energy power-law index $\alpha$. A certain sample of bursts
can thus be studied with the relations discussed above, to discern
the dominating radiation process. For instance, on the basis of
the  durations and hardnesses, bursts seem to  belong to two
distinct classes (e.g. \citet{B03}), or even three \citep{H98,
H04}. Any difference in radiation mechanism between these classes
can therefore be efficiently investigated.

\begin{figure}
\centering
\includegraphics[width=0.5\textwidth]{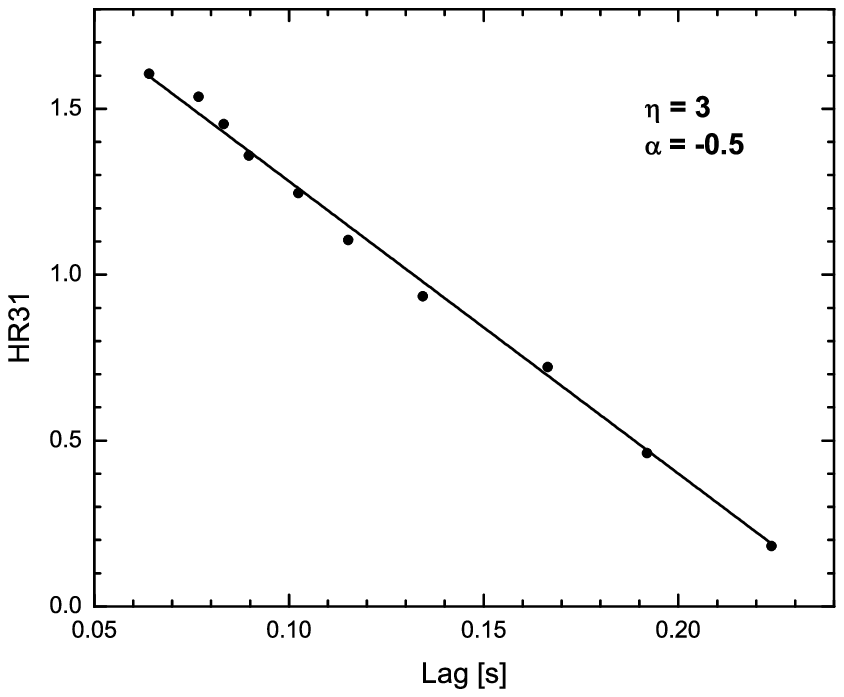}
\caption{Correlation between hardness ratio and spectral lag for a
"curvature-driven" pulse ($\eta = 3$). $\E00$ decreased to the
right. See the text for further details.
 \label{fig:HR4}}
 \end{figure}

It is also known that the importance of the curvature effect
varies between pulses (see e.g. \citet{RP02}). In the case of it
being important, the HIC index, $\eta$, will have a characteristic
behavior and a large value. In Fig. \ref{fig:HR}a, $\eta=3$ for a
curvature-driven pulse is marked by the dashed line and the
relation between HR31 and spectral lags for such pulses is shown
in Fig. \ref{fig:HR4}. This shows the hardness ratio decreasing
linearly with the corresponding time lag. In this particular run
$\alpha = -0.5$. $E_0$ decreases to the right with the point
furthest to the left with $E_0 = 1000$ keV and the point furthest
to the right with $E_0 = 100$ keV. This decrease is obvious in
Fig. 4, in which large values of $\eta$ and soft spectra sample a
space where the lag decreases with $E_0$. This, together with the
positive correlation between $\E00$ and HR31 depicted in Fig.
\ref{fig:HR3}, then gives the relation in Fig. \ref{fig:HR4}. On
the other hand, if the dominant time-scale is not the curvature
time-scale then the intrinsic $\eta$, representing the comoving
radiation process, will be revealed. For instance, assuming
synchtroton emission and a constant bulk Lorentz factor, $\Gamma$,
we expect $\Epk \propto \gamma^2_{min} \cdot B$ (the minimum
electron Lorentz factor and the magnetic field strength) and $F
\propto {\cal E}_e \cdot B^2\cdot \gamma_{min}$, where ${\cal
E}_e=\int_0^\infty \gamma N(\gamma)d\gamma$ is the total electron
energy ($N(\gamma)$ is the energy spectrum of electrons). If the
latter quantity is constant, we expect $\eta = 2$ if $B$ varies
and $\gamma_{min}$ is constant and $\eta = 1/2$ if $B$ is constant
and $\gamma_{min}$ varies. The observed correlations would then
sample different regions in the panels discussed above.

During the past few years increasing numbers of X-ray rich GRBs
and so-called X-ray flashes (XRF) have been observed (Heise et al.
2000, Barraud et al. 2003). Several studies have shown that the
main difference in properties between these populations and
classical GRBs is the peak energy distribution. It has therefore
been suggested that they are all of the same origin (e.g. Kippen
et al., 2002, 2003, Lamb et al., 2004) and that XRF is an
extension of the GRB energy distribution. As described above the
distribution of peak energy between different samples will have
characteristic signatures on the lag correlations and can be
further used in studying the difference between GRBs and XRFs.

\section{Conclusion}

The two spectroscopical analysis methods discussed in this paper,
HRS and LRS measure the same underlying behavior and are
alternative ways to represent the spectral evolution. We have
shown analytically and numerically that the spectral evolution
described by high spectral-resolution data, in empirical
correlations, naturally leads to correlations involving spectral
lags. The interpretation of the spectral lag is not
straightforward. This depends on the fact that the lag measures a
combination of spectral parameters such as $\eta$, $\E00$, and
$\tau$ (apart from effects caused by $\alpha$ and $\beta$; their
values and evolution). It further depends on the relative channel
widths into which the data are divided. For different pulses
within a burst, the spectral parameters are in general not the
same, and thus the lag is not expected to be the same during the
whole burst. This is indeed the case in observed bursts, as we
note in paper II (Ryde et al. 2004).

The main result of the above analysis is that the lag correlates
strongly with the decay time-scale of the pulse (or equivalently,
of the peak-energy decay) and that this relation is linear. This
means that the relationships including the lag, that have been
identified, should be translatable into relationships including
the pulse time-scale. This time-scale is closely connected to the
processes forming the pulse in the out-flowing plasma; the
radiation processes, acceleration mechanisms, light-travel effects
and outflow velocity etc. In particular, we find a good
correlation between lag and the quantity $F/\Phi_0$, reproducing
the observation by \citet{KL03}. We further find that harder
spectra, with large $\alpha$, have the largest spectral lags. This
is, in particular, the case for averaged $\eta$--values.
Similarly, an increase in $\E00$, increases the lag, except for
soft $\alpha$ (see Fig.4 d). Interestingly, the hardness ratio is
found to be only weakly dependent on $\eta$ and $\alpha$, and in
particular for low, initial peak energies $\E00$ it is practically
independent, and it is mainly determined by $\E00$; according to
Fig. 6a. The relation between the hardness ratio and the lags, for
a certain $\E00$ and $\eta$, are described by power-laws, as
$\alpha$ varies (Fig. 5d). Furthermore, for a curvature-driven
pulse ($\eta = 3$), the hardness ratio decreases linearly with lag
for lower and lower $\E00$.

The analytical and numerical results presented here are the
consequences of the empirical description of the spectral
evolution in pulses, which have been firmly established. These
results can thus be used as a reference in the analysis of
observed pulses and bursts, in particular low spectral-resolution
studies in which the spectral parameters cannot be found and
mainly the lag and hardness ratios are analyzed. Such an analysis
is performed in paper II (Ryde et al. 2004) in which we also
discuss how the conclusions presented here relate to the physical
models that have been proposed to explain the observed
lag-correlations.

\begin{acknowledgements}
I wish to thank Drs. A. M\'esz\'aros, S. Larsson and L. Borgonovo
for interesting discussions. The anonymous referee is also
acknowledged with thanks for providing suggestions that improved
the text. This study was supported by the Swedish Research
Council. Parts of the work was completed at the First Niels Bohr
Summer Institute at NORDITA, Copenhagen, Denmark.

\end{acknowledgements}

\end{document}